\documentclass[a4paper,10pt]{article}
\usepackage{amsmath}
\usepackage{amsfonts}
\usepackage{amssymb}

\usepackage{graphicx,epsfig}

\title{On the role of superconducting shields for propellantless propulsion}
\author{Luzi Bergamin and Dario Izzo}

\begin{document}

\maketitle

\begin{abstract}
We critically review our recent claims that it is possible to obtain a propellantless propulsion device similar to electrodynamic tethers by means of a closed wire partially shielded by a superconductor from the outer magnetic field. We find that such a device is not possible as it violates basic physical laws. Furthermore, we show what is the correct local picture for the currents distribution within the superconductor.
\end{abstract}
\section{Introduction}
There exist several contributions in the literature \cite{matloff1,matloff2,pinchook} where superconducting materials have been proposed as for an outer magnetic field onboard of a spacecraft. The concepts include the claims that superconducting materials can be used to shield forces and that a partially shielded current in this way could serve as an alternative to electrodynamic tethers. We critically review these concepts here and show that such a device cannot work.

The impossibility of such a device follows from a global evaluation of the Lorentz force acting on any finite and steady-current distribution $\vec J$ in a certain volume $V$: the net force due to a constant external magnetic field $B_0$ indeed is zero due to
\begin{equation}
 \vec F = \int_V dv \vec J \times \vec B_0 = \left( \int_V dv \vec J \right) \times \vec B_0 = 0\ ,
\end{equation}
where the current conservation $\vec \nabla \vec J = 0$ has been used and the result holds if now current flows through the surface $\partial V$.

Of course, local variations on the magnetic field may cause local forces, but these are just internal forces that do not generate a net force. This should also and in particular be true for the shielding by means of a superconductor unless fundamental properties of electrodynamics would be lost in that case.

\section{No external magnetic field}
\begin{figure}[t]
\centering \epsfig{file=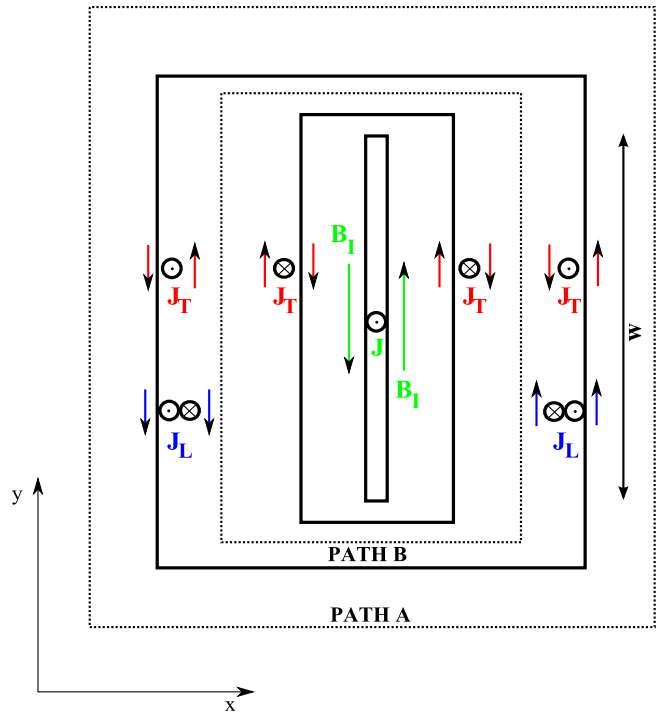,width=0.7\linewidth}
\caption{The situation of the superconducting shield if a current flows in the tape but no external magnetic field is present.}
\label{fig:1}
\end{figure}

Still, one would like to understand in detail why the concept cannot work. The aim of this short note is to discuss the forces that act on a closed wire in a magnetic field, if the former is partially shielded by a superconductor. In this section we consider the situation without an external magnetic field and ask what happens exactly in the superconductor. The situation we are considering is sketched in figure \ref{fig:1}: A current $I$ flows in a tape which is itself put into a hollow superconductor. The current is assumed to be closed somewhere far away from the superconductor such that it does not influence local physics. For simplicity we assume that the width $w$ of the tape (and therefore the y-extension of the superconductor) is much bigger than the thickness of the whole device in x direction. Therefore we may simply assume that all magnetic fields point in the y direction and are constant.

The current density $J=I/w$ in the tape produces a magnetic field $B_I = \mu_0 I/(2w)$. Can this be shielded by the superconductor? The answer follows from the integrated version of Amp\`ere's law if we take the area bounded by the path A in figure \ref{fig:1}. Then
\begin{equation}
\label{1.1}
 \oint_A \vec B(\vec r) d \vec r = \mu_0 \int_{\Sigma_A} \vec J (\vec r) \cdot \vec n (\vec r) d\sigma = \mu_0 I\ ,
 \end{equation}
where the result follows from the fact that there cannot flow any net currents in the superconductor (of course the latter is assumed to be of finite size.) It should be noted that the integration through the superconductor is harmless as long as one considers the magnetic induction $\vec B$, as this quantity is well defined everywhere. Though equation \eqref{1.1} tells us that the magnetic field effectively cannot be shielded outside of the superconductor, it does not yet say how the latter is produced there. Indeed, the current $I$ flowing in the tape cannot be responsible, as its magnetic field \emph{is} shielded by the superconductor. For each point on the path A there exists a region between that point and the tape, where $\vec B \equiv 0$. Therefore $\vec B$ outside of the superconductor is produced by surface currents that can flow in the London layers of the outer surface of the superconductor.

Let us consider these currents a little bit more in detail. The superconductor is assumed to be thick, i.e.\ the thickness is much bigger than the London penetration depth, so there exists a region far enough from the surface where $\vec B \equiv 0$. The currents effectively shielding the magnetic field flow at the surface in the so-called London layers. A superconductor has magnetic susceptibility $\chi_m=-1$, so the split of the magnetic induction as $\vec B = \mu_0 (\vec H + \vec M)$ is not defined uniquely. As the superconductor considered here is not simply connected there exist two different types of currents; the first type closes topologically non-trivial around the region with $\vec B \equiv 0$ and is denoted by $I_T$ (cf.\ also figure \ref{fig:1}), the second type closes topologically trivial (i.e. local within one London layer) and therefore reminds of the characteristic of magnetization, it will be referred to as $I_L$. Although these namings are arbitrary, the physical nature of the two different types of currents is essential. Which type of current is responsible to shield the magnetic field produced by the tape in the superconductor? To answer this question we again use the integrated version of Amp\`ere's law, but this time we integrate around path B in figure \ref{fig:1}. In this way we obtain
\begin{equation}
\label{1.2}
 \oint_B \vec B(\vec r) d \vec r = 0 =  \mu_0 \int_{\Sigma_B} \vec J (\vec r) \cdot \vec n (\vec r) d\sigma\ ,
 \end{equation}
so the total current flowing through the area must vanish. The topologically trivial currents clearly cannot contribute so equation \eqref{1.2} tells us that
\begin{equation}
 I_T = -I/2\ .
\end{equation}
Therefore in both inner surface layer there flows a current with strength $-I/2$ and consequently the opposite currents flow in the outer surface layers. In the current situation no further shielding is necessary at the inner surface. Consider e.g.\ the inner right wall: adding together the magnetic field from the current $I$ and from the current $I_T$ at the inner left wall yields a total magnetic field $B_I/2$ to be shielded, which is exactly balanced by the current $I_T$. This last conclusion is a consequence of the particular geometry chosen and will not generalize to different situations.

Having concluded what happens at the inner surface, let us now consider the outer one. The topologically non-trivial current $I_T$ of course is of the same strength as at the inner surface. It produces a magnetic field $B_T = \mu_0 I/(4w)$ as indicated in figure \ref{fig:1}. Outside of the superconductor this is half of the magnetic field needed to satisfy Amp\`ere's law in equation \eqref{1.1}, at the same time $I_T$ causes a magnetic field inside the superconductor, which must be shielded by means of local currents or magnetization, resp. As the difference in the magnetic field between the surface and the center of the superconductor has the correct value solely with the current $I_T$,
\begin{equation}
B_T(\mbox{\small surface}) - B_T(\mbox{\small center}) = \frac{\mu_0 I}{2w}\ ,
\end{equation} 
the local currents $I_L$ must induce a constant shift of the magnetic field by $\mu_0 I/(4w)$ in such a way that the correct boundary conditions on the magnetic field are met. In terms of a magnetization $\vec M$ this would correspond to the constant case $M(\mbox{\small surface})= M(\mbox{\small center}) \neq 0$.
This terminates the calculation of the case without external magnetic field.
\section{External magnetic field}
\begin{figure}[t]
\centering \epsfig{file=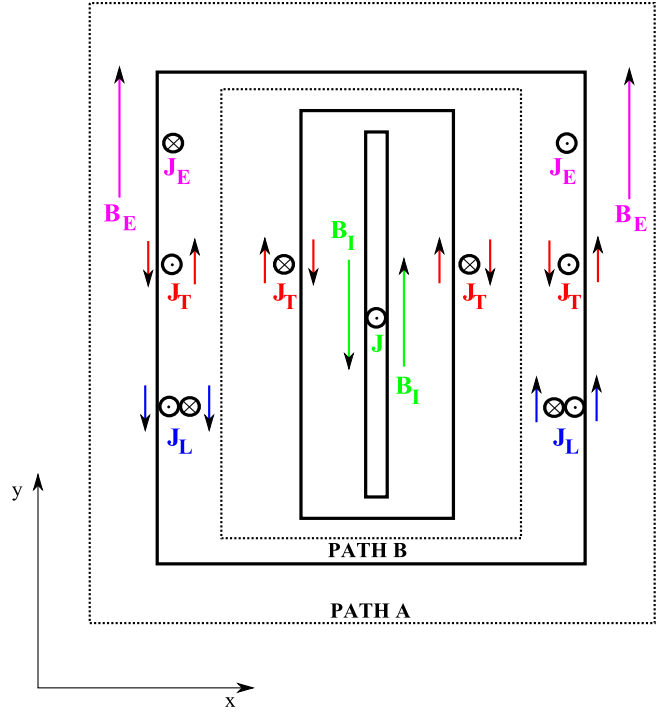,width=0.7\linewidth}
\caption{The situation of the superconducting shield if a current flows in the tape and external magnetic field is present.}
\label{fig:2}
\end{figure}

Now we slightly modify the situation by turning on an external magnetic field $B_E$ (treated as a background field.) The new situation is depicted in figure \ref{fig:2}.

As is easily seen this external magnetic field can be shielded by a current flowing around the complete superconductor (instead of just one wall as in the case of $I_T$.) In order to prevent an overloaded picture no arrows for the induced magnetic fields due to $I_E$ have been added. What is the total force acting on the superconductor? The relevant expression for the Lorentz force reads
\begin{equation}
 \vec F = l \vec I \times \vec B\ ,
\end{equation}
where $l$ is the length over which the current flows. In our case the magnetic field is the external field $B_E$. The current is $I_T$ flowing in the London layer of the outer surface. The interaction can be made explicit by taking the integral over the complete London layer:
\begin{equation}
 F_x = - A \int_S^0 J_z B_y dx = - \frac{A}{\mu_0} \int_S^0 \partial_x B_y B_y = \left.\frac{1}{2} B_y^2 \right|_S^0\ ,
\end{equation}
where $A$ denotes the area of one of the outer walls of the superconductor and the integral is taken from the surface $S$ to the center $0$ of the superconducting wall.
Adding both sides and carefully choosing the sign one obtains
\begin{equation}
\label{finalresult}
 F_x = - \frac{A}{\mu_0} \left( (B_E + B_I)^2 - (B_E-B_I)^2 \right) = - l B_E (\frac{2w B_I}{\mu_0}) = -l B_E I\ .
\end{equation}
Therefore it is seen that the same Lorentz force acts on the shield that would act on the wire and therefore the shielding by a superconducting cylinder does not work. This now is intuitively clear as on both surfaces together there flows a total current $I$.

\section{Conclusions}
From the force in equation \eqref{finalresult} acting on the superconducting shield it is seen that the net Lorentz force acting on a superconducting shield is exactly the same as the one that would act on a bare wire. Therefore a propellantless propulsion device relying on the concept of a partially shielded wire cannot work. The same conclusion is held by observing simply that on any current cloud there can be no net force if this is immersed on a constant magnetic field.

\subsection*{Acknowledgement}
The authors would like to thank J.R.~Sanmartin and E.C.~Lorenzini who had the good will of prooving them wrong.

\bibliographystyle{fullsort}
\bibliography{BeIzPi}

\end{document}